# Incremental Learning with Accuracy Prediction of Social and Individual Properties from Mobile-Phone Data

Yaniv Altshuler, Nadav Aharony, Alex ("Sandy") Pentland
MIT Media Lab,
Cambridge, MA, 02139
{yanival, nadav, sandy}@media.mit.edu

Michael Fire, Yuval Elovici
Deutsche Telekom Laboratories at Ben-Gurion University of the Negev,
Department of Information Systems Engineering, Ben Gurion University of the Negev
Be`er-Sheva 84105, Israel
{mickyfi, elovici}@bgu.ac.il

## ABSTRACT

As truly ubiquitous wearable computers, mobile phones are quickly becoming the primary source for social, behavioral, and environmental sensing and data collection. Today's smartphones are equipped with increasingly more sensors and accessible data types that enable the collection of literally dozens of signals related to the phone, its user, and its environment. A great deal of research effort in academia and industry is put into mining this raw data for higher level sense-making, such as understanding user context, inferring social networks, learning individual features, predicting outcomes, and so on. In many cases, this analysis work is the result of exploratory forays and trial-and-error. Adding to the challenge, the devices themselves are a limited platform, and any data collection campaign must be carefully designed in order to collect the right signals, in the appropriate frequency, and at the same time not exhausting the device's limited battery and processing power. There is need for a more structured methodology and tools to help with designing mobile data collection and analysis initiative.

In this work we investigate the properties of learning and inference of real world data collected via mobile phones over time. In particular, we look at the dynamic learning process over time, and how the ability to predict individual parameters and social links is incrementally enhanced with the accumulation of additional data. To do this, we use the Friends and Family dataset, which contains rich data signals gathered from the smartphones of 140 adult members of a young-family residential community for over a year, and is one of the most comprehensive mobile phone datasets gathered in academia to date.

We develop several models that predict social and individual properties from sensed mobile phone data, including detection of life-partners, ethnicity, and whether a person is a student or not. Then, for this set of diverse learning tasks, we investigate how the prediction accuracy evolves over time, as new data is collected. Finally, based on gained insights, we propose a method for advance prediction of the maximal learning accuracy possible for the learning task at hand, based on an initial set of measurements. This has practical implications, like informing the design of mobile data collection campaigns, or evaluating analysis strategies.

## Keywords
Rich data, reality mining, social network, mobile sensing, inferring attributes

## 1. INTRODUCTION
Mobile phones, and increasingly smartphones, have become an integral part of many people's everyday lives. Users carry their smartphone almost everywhere, and use it in order to perform many of their day-to-day communication and activities. These include connecting with family and friends via voice calls or text messaging, searching for information on the Internet, installing and using different mobile applications for business and pleasure, using location based services such as navigation instructions, or using the smartphone as alarm clock in order to wake up on time in the morning.

The pervasiveness of mobile phones has made them popular scientific data collection tools, as social and behavioral sensors of location, proximity, communications and context. Eagle and Pentland [1] coined the term ``Reality Mining'' to describe collection of sensor data pertaining to human social behavior. While existing works have demonstrated results for modeling and inference of social network structure and personal information out of mobile phone data, most are still mainly proofs of concept in a nascent field. The work of the "*data scientist*" is still that of an *artisan*, using personal experience, insight, and sometimes "gut feeling", in order to extract meaning out of the plethora of data and noise.

As the field of computational social science matures, there is need for more structured methodology. One that would assist the researcher or practitioner in designing data collection campaigns, understanding the potential of collected datasets, and estimating the accuracy limits of current analysis strategy vs. alternative ones Such methodology would assist it in the process of maturing from a field of "craft" into a field of science and engineering.

In this work, we present a first step in this direction. Specifically, we investigate the learning and prediction of social and individual models from raw phone-sensed data. We focus on social ties and individual descriptors that can be tied to social affiliation and affinity. For these prediction tasks, we look at the dynamic learning process over time, and how the ability to predict individual parameters and social links is enhanced over time with the accumulation of additional data.

To do this, we use the *Friends and Family* dataset, which contains rich data signals gathered from the smartphones of 140 adult members of a young-family residential community for over a year[2], as well as self-reported personal and social-tie information.

We first build classifiers for predicting personal properties like nationality or gender. We then proceed to predict more complicated social links such as the subject's life-partner, or "significant other".

When analyzing the improvement in performance of the social prediction over time, we show that it resembles the *Gompertz* function – a known mathematical model that has been used to approximate many processes in a variety of fields, including growth of tumors and adoption of technological services in communities, among others.

Our key contributions presented in this paper are as follows:

- We demonstrate characteristics of incremental learning of multiple social and individual properties from raw sensing data collected from mobile phones, as the information is accumulated over time.
- We show that for different learning tasks, prediction methods, and input signals, the evolving learning of social and individual features, as mobile phone sensing data accumulates over time, can be fitted to the form of a *Gompertz* function.
- Furthermore, we propose a method for advance prediction of the maximal learning accuracy possible for the learning task at hand, using just the first few measurements. This information can be useful in several ways, including:
    - Informing real-time resource allocation for data collection, for an ongoing data collection campaign.
    - Estimating accuracy limits and time needed for desired accuracy level of a given method.
    - Early evaluation of modeling and learning strategies.
- Finally, we present new models for predicting social and individual features from raw mobile-phone sensed data, which were developed as part of the methodological analysis.

The paper is organized as follows: We start by presenting related work in Section 2. In section 3 we discuss the methodology of the experiment and our learning techniques. Section 4 contains the results, and discussion and future work appear in Section 5. Concluding remarks are given in Section 6.

## 2. SCIENTIFC BACKGROUND

In recent years, the social sciences have been undergoing a digital revolution, heralded by the emerging field of ''computational social science''. Lazer, Pentland, et al. [3] describe the potential of computational social science to increase our knowledge of individuals, groups, and societies, with an unprecedented breadth, depth, and scale. Computational social science combines the leading techniques from network science [4-6] with new machine learning and pattern recognition tools specialized for the understanding of people's behavior and social interactions [7].

### 2.1 Mobile Phones As Social Sensors

The pervasiveness of mobile phones the world over has made them a premier data collection tool of choice, and they are increasingly used as social and behavioral sensors of location, proximity, communications and context. Eagle and Pentland[1] coined the term "*Reality Mining*" to describe collection of sensor data pertaining to human social behavior. They show that using call records, cellular-tower IDs, and Bluetooth proximity logs, collected via mobile phones at the individual level, the subjects' regular patterns in daily activity can be accurately detected[1, 7]. Furthermore, mobile phone records from telecommunications companies have proven to be quite valuable for uncovering human level insights. As one example, Gonzales et al. show that cell-tower location information can be used to characterize human mobility and that humans follow simple reproducible mobility patterns[8]. This approach has already expanded beyond academia, as companies like Sense Networks [9], are putting such tools to use in the commercial world to understand customer churn, enhance targeted advertisements, and offer improved personalization and other services.

### 2.2 Individual Based Data Collection

On one hand, data gathered through service providers include information on very large numbers of subjects, but on the other hand, this information is constrained to a specific domain (email messages, financial transactions, etc.), and there is very little if any contextual information on the subjects themselves. The alternative approach, of gathering data at the individual level, allows collecting many more dimensions related to the end user, many times not available at the operator level. Madan et al.[10], follow up on Eagle and Pentland's work [1], and show that mobile social sensing can be used for measuring and predicting the health status of individuals based on mobility and communication patterns. They also investigate the spread of political opinion within a community [11]. Other examples for using mobile phones for individual-based social sensing are those by Montoliu et al. [12], Lu et al. [13], and projects coming from CENS center, e.g. Campaignr by Joki et al. [14], and additional works as described in [15]. Finally, the Friends and Family study, which our paper uses as its data source, is probably the richest mobile phone data collection initiative to date as the number of signals collected, study duration, and the number of subjects. The technical advancements in mobile phone platforms and the availability of mobile software development kits (SDKs) to any developer is making the collection of Reality Mining type of data easier than ever before.

In addition to mobile phones, there have been other types of wearable sensor-based social data collection initiatives. A notable example is the *Sociometric Badge* by Olguin et al. which captures human activity and socialization patterns via a wearable sensor badge and are used mostly for data collection in organizational settings [16]. The results of our work are applicable to these types of studies as well.

### 2.3 Learning and Prediction of Social and Individual Information

Many studies involving predicting individual traits and social ties were conducted in the recent years in the general context of social networking. As few examples, relevant works have been published by Liben-Nowell and Kleinberg [17], Mislove [18] ,and Rokach et. al. [19]. These works combine machine learning algorithms together with social network data in order to build classifiers.

# 3. METHODOLOGY
## 3.1 Mobile Data Collection System

Aharony et al.[2]. developed a social and behavioral sensing platform that runs on Android operating-system based mobile devices, which can continuously record a broad range of data signals. Each type of signal collected by the system is encapsulated as a conceptual ''probe'' object. The probes terminology is used rather than ''sensors'' as probes include traditional sensors such as GPS or accelerometer, but also other types of information not traditionally considered as sensor data, like file system scans or logging user behavior inside applications. Additional signals include information such as cell tower ID, wireless LAN IDs; proximity to nearby phones and other Bluetooth devices; call and SMS logs; statistics on installed phone applications, running applications, media files, general phone usage; and other accessible information.

The dataset described in the next section was collected using this system, with a configuration that included over 25 different types of data signals. The deployment also included an on-phone survey component, and integrated applications such as an alarm clock app. Figure 1 illustrates the deployed system configuration, enabling automated data upload, as well as remote configuration settings and remote updating of the system itself. Figure 2 gives an overview of the back-end side of the system. The software system, named "*Funf*", has been released as open source and available at [20].

The "Friends and Family" living laboratory study was conducted over a period of 15 months between March 2010 and June 2011, with a subject pool of 140 individuals. It is the first study conducted under the Social fMRI methodology, which uses mobile phones together with a data-rich collection approach to create a "virtual imaging chamber" around a community *in-situ* [2]. To the best of our knowledge, it is the most comprehensive mobile phone experiment performed in academia to date.

### *3.1.1 Community Overview*
The research goals of the Friends and Family study touch on many aspects of life, from better understanding of social dynamics to health to purchasing behavior to community organization. It was conducted with members of young-family residential living community adjacent to MIT. All members of the community are couples, and at least one of the members is affiliated with the university. The community is composed of over 400 residents, approximately half of which have children. In March 2010 the first pilot phase of the study was launched with 55 participants, and in September 2010, the second phase of this study was launched with 85 additional participants. The participants were selected randomly, in a way that would achieve a representative sample of the community and sub-communities.

### *3.1.2 Privacy Considerations*
The study was approved by the Institutional Review Board (IRB) and conducted under strict protocol guidelines. One of the key concerns in the design of the study was the protection of participant privacy and sensitive information. For example, data is linked to coded identifiers for participants and not their real world personal identifiers. All human-readable text, like phone numbers and text messages are captured as hashed identifiers, and never saved in clear text. Collected data is physically secured and de-identified before being used for aggregate analysis.

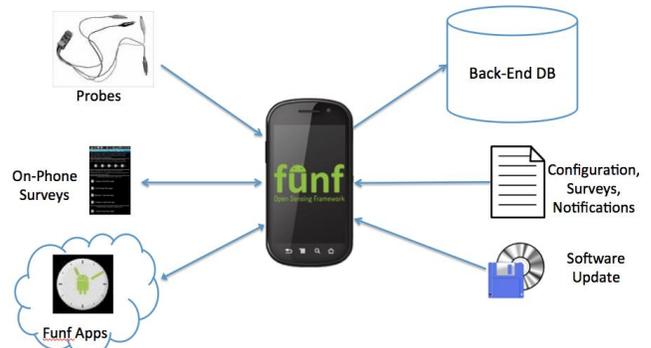

**Figure 1. Friends and Family Phone System Overview**

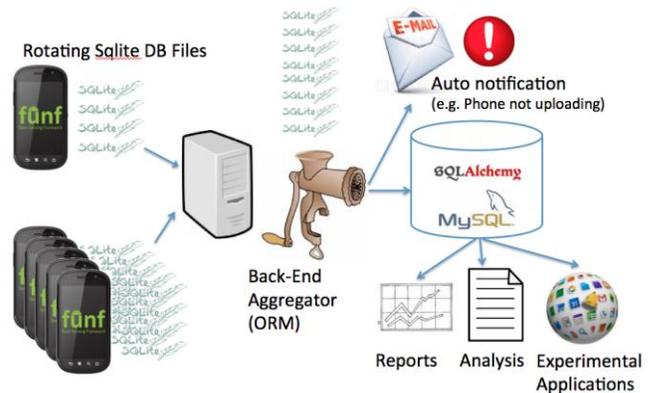

**Figure 2. Back-End Data Aggregation Overview**

### *3.1.3 Friends and Family Dataset*
To the best of our knowledge, the dataset generated from the study is probably the largest and richest dataset ever collected on a residential community to date. The accumulated size of the database files uploaded from the study phone devices adds up to *over 60 Gigabytes.* The data is composed of over *30 million* individual scan events (for all signals combined), where some events capture multiple data signals. Just as example, the dataset includes:

- 20 million wifi scans, which in turn accumulated 243 million total scanned device records.
- 5 million Bluetooth proximity scans, which in turn accumulated 16 million total scanned device records.
- 200,000 phone calls.
- 100,000 text messages (SMS).

In the analysis presented in this paper we give special focus to the data that was collected in November 2010 and April 2011, after the mobile platform was improved, new features such as different call types where added, and several hardware problems where fixed. These two months are ones where there were no major holiday breaks in the academic schedule of the university, and the bulk of participants were physically on campus.

In addition to the phone-based data, the study also collected personal information on each participant The dataset includes information on age, gender, religion, origin, current and previous income status, ethnicity, and marriage information, among others.

## 3.2 Machine Learning Predictions

In order to evaluate learning over time, which is the main goal of our current work, we needed a set of learning and prediction models to work with. These are mostly illustrative models, which enable us to conduct our main analysis.

In order to achieve our final goal of predicting participants' personal and social information, we utilized two approaches – first is a machine learning approach, described in this section, and the second is a social network based prediction approach, described in the following section.

The first step in applying the machine learning methodology is to create feature vectors for each participant in the study. Each feature vector contains information on the participant's communication and phone usage patterns as were collected during the study.

In order to cope with the huge amount of data collected during the study, we developed code using C# and *Python's NetworkX* library [21]. Our code parsed the collected data, and extracted feature vectors for each participant. We extracted 32 different features within a specified time interval. Namely, we collected the following features for each participant:

- **Internet usage features**: we calculate the number of distinct searches performed using the phone's browser, and the number distinct bookmarks saved by the user.
- **Calls pattern features**: we compute the total number of calls, the number of unique phone numbers each user was in contact with, and the total duration of all calls. We also calculate the number of incoming/outgoing/missing calls and the total call durations according per call type.
- **SMS messages pattern features**: we compute the total number of SMS messages, the number of unique phone numbers each participant connected with via SMS, and the of total incoming/outgoing SMS messages.
- **Phone applications related features:** we count the number of applications installed and uninstalled on each device. We also compute the total number of currently running applications (originally sampled every 30 seconds).
- **Alarm features:** we count the number of alarm-clock alarms and the number of "snooze" presses for each participant that used our alarm clock app.
- **Location features:** we calculated the number of different cellular cell tower ids and the number different wifi network names names seen by the smartphone**.** These features act as a rough indication of the number of different locations a participant visited during the time period.

Our next step was to extract all participant features for different time intervals. Using the extracted features we can build different classifiers that are able to predict the participants` personal information. We used the *WEKA* software [22] in order to test different machine learning algorithms. In our experiments we evaluated a number of popular learning methods: we used *WEKA*'s C4.5 decision trees, Naive-Bayes, Rotation-Forest, Random-Forest, and AdaBoostM1. Each classifier was evaluated using the 10-fold cross validation approach, and in order to compare results between different classification algorithms, we used each classifier's Area Under Curve, or AUC measure (also referred to as ROC Area) and F-measure results. In order to obtain an indication of the usefulness of various features, we analyzed their importance using *WEKA*'s information gain attribute selection algorithm

Using the machine learning approach we built five different classifiers that predict the following: (1) the gender of the participant, (2) whether the participant is a student or not, (3) whether the participant has children or not, (4) whether the participant is above the age of 30, and (5) whether the participant is a native US citizen or not.

## 3.3 Social Network Predictions

Another method for predicting a participant's personal information details is using the participants' different social networks. Using the data collected in the study. We can span different types of social networks between the participants, according to different interaction modalities. Namely, we can define the following social networks:

- **SMS Social Network**: we can construct the community's SMS messages social network (see Figure 2) as a weighted graph $G_s = <V_s, E_s>$ according to the SMS messages the participants sent. Each weighted link $e = (u, v, w) \in E_s$ in this social network represent connection between two different phone numbers $u, v \in V$, while *w* is the strength of the link defined as the number of SMS message send between the two phone number[1]. The SMS network also includes encoded phone numbers outside of the study which were contacted by more than one participant.
- **Bluetooth Social Network:** we can construct a weighted network graph $G_s = <V_B, E_B>$ of face-to-face interaction according to information collected about nearby Bluetooth devices. Each link $(u, v, w) \in E_s$ in this social network represent the fact that the two devices $u, v \in V_B$ encounter each other at least one time, while the *w* is the strength of the link, defined as the number of times the two devices encounter one another.
- **Calls Social Network:** Similar to the SMS social network, we can construct a network based on the participant's call graph $G_C = <V_C, E_C>$ according to the participants` phone calls. In this social network each link $(u, v, w) \in E_C$ represents the fact at least one call was made between two different phone numbers ,$u, v \in V_C$, while *w* is the strength of the link. defined as the number of calls between *u and v*.

By using the social networks defined above, together with different graph theory algorithms, we can predict different types of personal and social information. In order to predict the participants' significant other we analyzed the Bluetooth social network. We predicted that each participant's significant other is the person that the participant spent the most time with during the measured interval. Namely, let $u \in V_B$ then:

$$significant\text{-}other(u)) = \{v | (u, v, w) \in E_B \text{ and}$$
$$\forall (u`, v`, w`) \in E_b \ w > w`\}$$

---

[1] In some cases, the number interaction may not be fully accurate due to the fact we do not have the full connection information for phone number outside the study

In order to predict the subjects' ethnicity we used the SMS social network (Figure 3). We used the Louvain algorithm for community detection [23], which separates the graph into disjoint groups.

At each iteration, we assume that we have information on the ethnicity of at least some of the nodes. The general idea is to then generate an ethnicity prediction for the members of each detected community based on the ethnicity of the majority of known nodes in that community. This is similar to the ideas of the label propagation approach [24] and in [18].

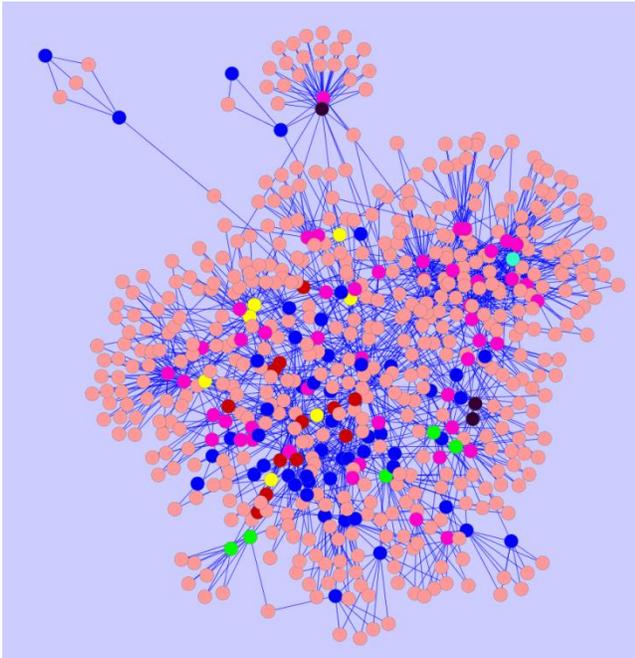

**Figure 3. SMS Social Network Graph of created over 65 weeks (graph also includes unknown out-of-study nodes, which connect to at least two known in-study nodes). Different vertex colors represent different ethnicity[2].**

### 3.4 Prediction Accuracy Evolution over Time

As discussed, the goal of this work is to study and analyze the evolution of the learning process of personal features and behavioral properties along the time axis. For this analysis, we care less about the specific learned models and their generalizability, but rather care about using them to study and benchmark the evolution of the learning process as data accumulates. Understanding this process is of significant importance to researchers in a variety of fields, as it would provide approximation for the amount of time that is needed in order to "learn" these features for some given accuracy, or alternatively, what is the level of accuracy that can be obtained for a given duration of time.

In order to model this process we used the *Gompertz function*:

$$f = ae^{be^{ct}}$$

---

[2] All graphs in this paper where created by using *Cytoscape* software

This model is flexible enough to fit various social learning mechanisms, while providing the following important features:

(a) Sigmoidal advancement, namely – the longer the process continues the more precise its conclusions will be.

(b) The rate at which information is gathered is smallest at the start and end of the learning process.

(c) Asymmetry of the asymptotes, implied from the fact that for any value of *t*, the amount of information gathered in the first *t* time steps is greater than the amount of information gathered at the last *t* time steps.

The *Gompertz function* is frequently used for modeling a great variety of processes (due to the flexible way it can be manipulated using the parameters *a*, *b,* and *c*), such as mobile phone uptake [25], population expansion in a confined space [26], or growth of tumors [27]

Following is an illustration of the *Gompertz function:*

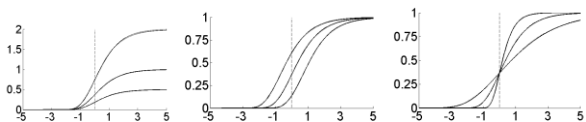

**Figure 4. An illustration of the *Gompertz function*. The charts represent the following functions (from left to right):** $y = xe^{-e^t}$, $y = e^{-xe^t}$ and $y = e^{e^{-xt}}$, for $x = \frac{1}{2}$, $x = 1$ and $x = 2$.

The applicability of the *Gompertz function* for the purpose of modeling the evolution of locally "learning" the preferences and behavior patterns of users was demonstrated in [28], where a prediction of the applications that mobile users would chose to install on their phones was generated using an ongoing learning process, and closely resembles the form of the Gompertz function.

For generating the models presented here, we have ran a Gompertz regression on data obtained from the predictors and classifiers developed using the methods described above. Each predictor/classifier was executed on data gathered between November 1th and November 30th, 2010. Starting from an input of a single day (November 1[st]), in each consecutive execution, another day of data was added to the input (so that iteration #1 was on data from November 1[st], execution #2 had input of data two days, November 1 and 2 together, and so on, until an accumulation of 30 days in which the classifier ran on data from the entire month of November. Figure 7 - Figure 10 in the results section present the results of 4 of the classifiers that we have run (more results were omitted due to space considerations and will appear in an extended version this work).

## 4. RESULTS

### 4.1 Machine Learning Classifiers Results

Using the machine learning algorithms we succeed in predicting different personal information. Our prediction results vary according to the amount of data, the number of features, and the time periods for which the classifier ran on.

- **Gender prediction** - we predicted the gender of the participants. Our dataset included the gender information on 103 participants. Our decision tree classifier (J48) got AUC of 0.642 and F-measure of 0.611, where the most influential features where the number of Internet searches and the number of alarms. In general, female participants perform fewer search queries using their smartphones.
- **US-natives prediction** – we tried to predict whether the origin of the participant is inside or outside the United States. Our dataset contained information about the origin of 86 participants. Our Naïve-Bayes classifier got AUC of 0.728 and an F-measure of 0.806. Where the most influence features where: the number of incoming and outgoing SMS. In general, participants born outside the United States send and receive fewer SMS messages than US natives.
- **Have children prediction** – we tried to predict which of the participants in the study have children. Our dataset contained information about the children of 63 participants. Our Naïve-Bayes classifier got AUC of 0.803 and an F-measure of 0.682, when using only four features: Number of missing calls, total number of application installed, distinct number of application installed, and number of alarms set. In general, participants that have children have more missed calls and fewer applications installed.
- **Is student prediction** – we tried to predict which of the study participants are students (vs a different occupation). Our dataset contained information on about 88 participants, almost half of them are students. Our Rotation-Forest classifier gave AUC of 0.639 and an F-measure of 0.625
- **Age prediction** - we tried to predict which of the study participants are above 30 years old or above. Our dataset contained information about 80 participants, out of them 34 were age 30 or above. Our decision tree classifier (J48) got AUC of 0.592 and an F-measure of 0.562, where the most influential features where the number of Internet searches and the number of calls. In general, participants above the age of 30 performed fewer search queries using their smartphones.

**Table 1. Predicting Personal Information Results**

|  | Influential Features | AUC | F-Measure |
|---|---|---|---|
| Age | Searches Number | 0.592 | 0.562 |
| Children | Missing Calls | 0.803 | 0.682 |
| Gender | Searches Number | 0.642 | 0.611 |
| Student | - | 0.606 | 0.608 |
| Origin | SMS Messages | 0.728 | 0.806 |

### 4.2 Social Network Predictions Results
We predicted that each participant's significant other is the person that the participant spent the maximum time with during the study according to the Bluetooth social network graph. We ran this prediction on the face-to-face interactions Bluetooth graph that was created during time period of 30 days in November 2010 (Figure 3). Our prediction succeeded in classifying 65.6% of the couples (44 out of 67).

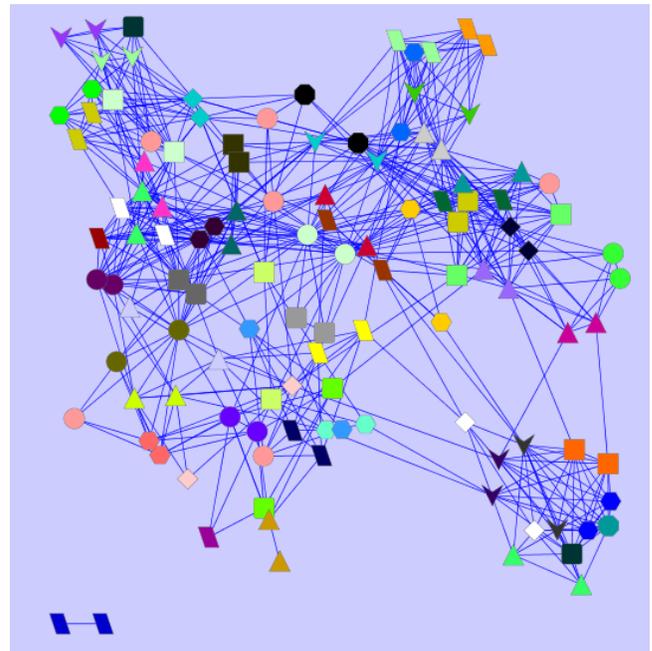

**Figure 5. Bluetooth social network graph of face-to-face interaction during November 2010– significant other are with the same shape and color each link represents at least 100 interaction.**

The Louvain method for community detection partitioned the SMS social network into 13 disjoint groups (Figure 5). Using our method we succeeded in predicting the ethnicity of 60% of the participants (77 out of 128).

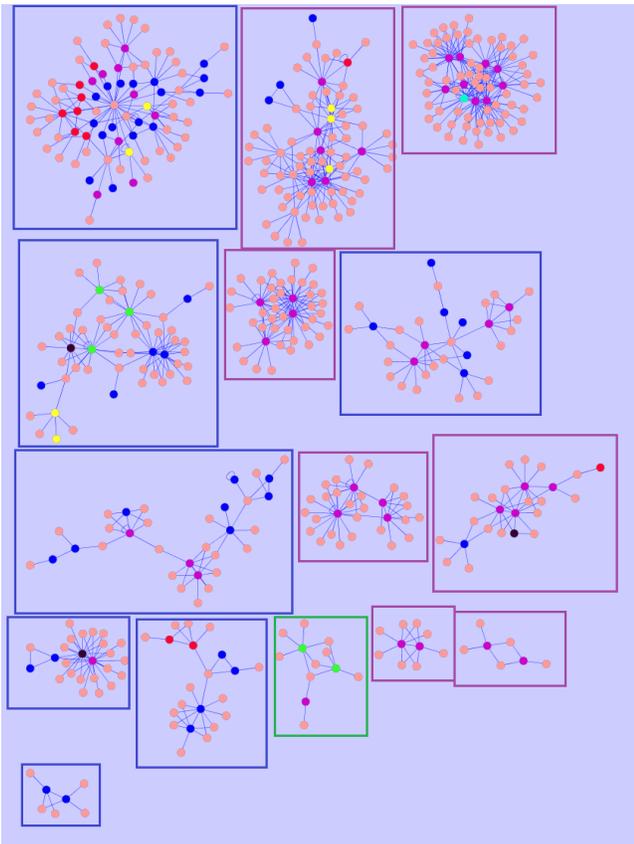

**Figure 6. Partitioned SMS Social Network Using Louvain Algorithm–each group have different ethnicity according to the major ethnicity of the group (similar to label propagation algorithms) (Blue: Asian, Purple: White, Green: Middle Eastern)**

## 4.3 Incremental Learning over Time Results

We have shown in the previous subsections that using different communication patterns and social network graphs we are able to predict specific personal and social information. Our next step was to examine how our classifiers evolve over time. We ran different classifiers with increasingly accumulating daily data that was collected from the month of November 2010. We obtained the following results for four of the classifiers, as presented in Figure 7 - Figure 10.

Figure 7 shows the classifier for whether a participant is US born or not (e.g. an international student or their spouse). The vertical axis represents the area under curve (AUC) values. The fitted Gompertz function has parameters of (0.8, -0.4, -0.14), with regression residual standard error of 0.02591, and achieved convergence tolerance of 7.404e-06.

Figure 8 shows the classifier for whether a participant is a student or not. Again, the vertical axis represents AUC, values. The fitted Gompertz function has parameters of (0.69, -0.35, -0.06), with regression residual standard error of 0.02237, and achieved convergence tolerance of 4.095e-06.

Figure 9 shows the classifier for whether we can predict that a participant's significant other. The vertical axis represents the percentage of correct matches. The fitted Gompertz function has parameters of (0.66, -0.78, -0.12), with regression residual standard error of 0.02762, and achieved convergence tolerance of 1.505e-06.

Figure 10 shows the classifier for whether we can predict a participant's ethnicity. The vertical axis represents the percentage of correct predictions. The fitted *Gompertz* function has parameters of (0.68, -2.18, -0.05), with regression residual standard error of 0.06676, and achieved convergence tolerance of 5.568e-06.

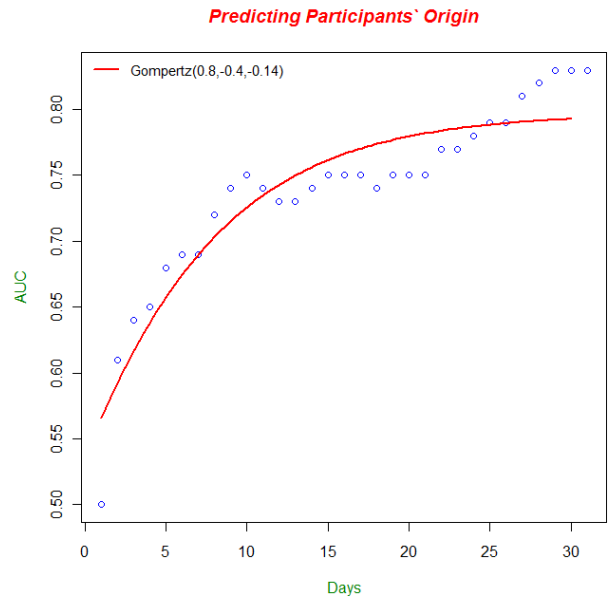

**Figure 7. Participants' origin Naïve-Bayes classifiers AUC results**

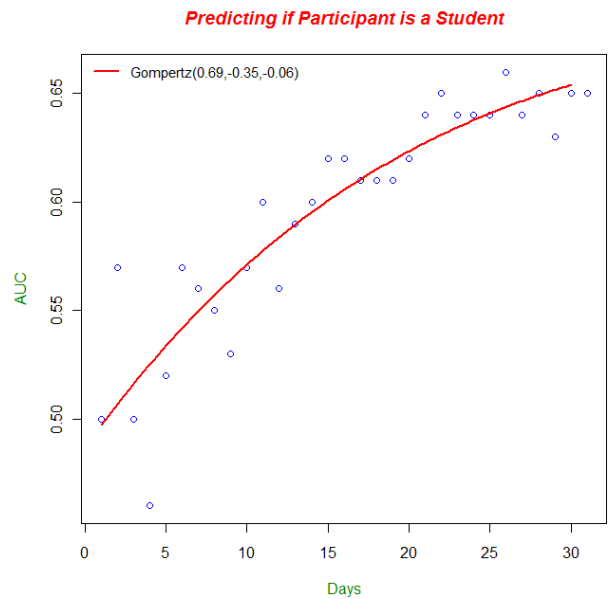

**Figure 8. Predicting If the Participant is a Student over Time: Rotation-Forest Classifier AUC results**

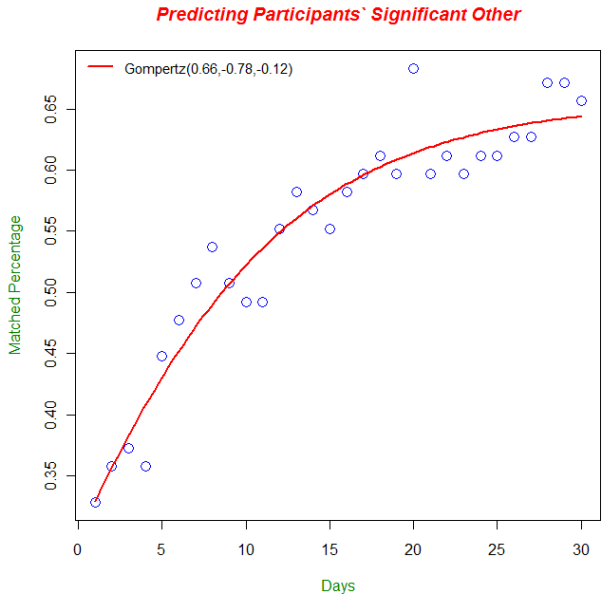

**Figure 9. Predicting Significant Other over Time – we chose the significant other as the node with the maximum strength.**

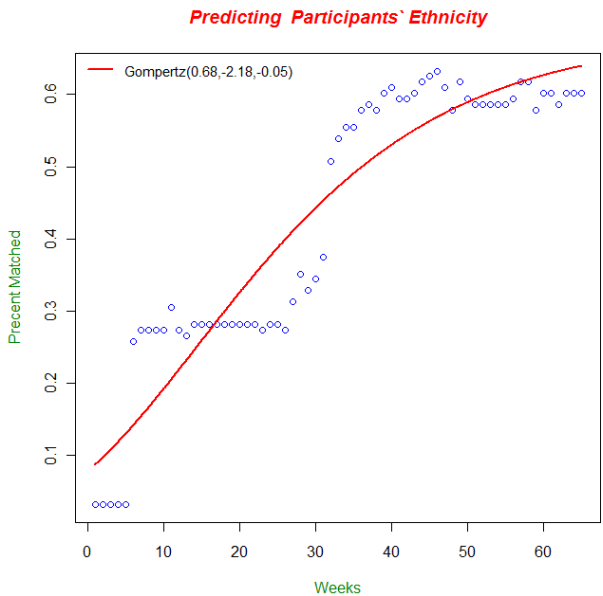

**Figure 10. Predicting ethnicity using SMS social network over time (65 weeks) – after every week we analyze the graph with the same method as described at 3.4 (Louvain Algorithm).**

Figure 11 demonstrates the correlations among the learning process dynamics of several features. It was calculated using the Pearson product-moment correlation coefficient (a measure of the linear dependence between two variables X and Y, giving a value between +1 and −1). The correlation is defined as the covariance of the two variables divided by the product of their standard deviations. In general, variables of correlation higher than 0.5 are usually considered strongly correlated.

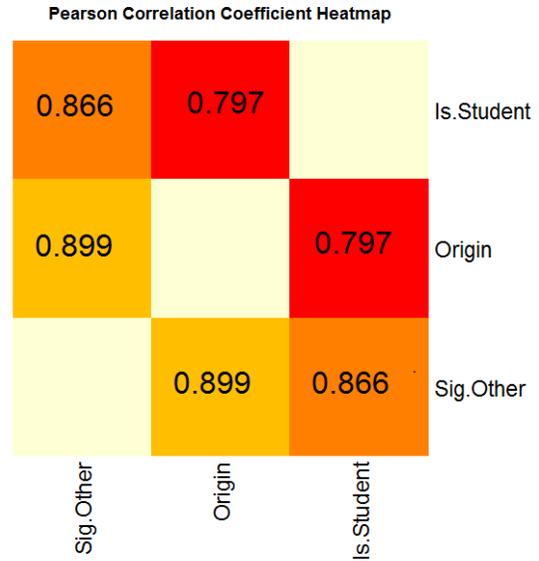

**Figure 11.** Pearson correlation between the learning process dynamics for three of the properties we predict. As might be expected, there are some strong correlation between the different evolution trajectories of the learning processes of the three features. However, notice that while some are very highly correlated (e.g. Origin \ Significant other), which might point out a strong correlation in the underlying data itself (i.e. people tend to get married more within the same ethnic group), other display lower correlation (e.g. Origin \ Is student).

## 5. DISCUSSION AND FUTURE WORK

As reviewed in section 3.4, the *Gompertz* function is a well-known technique that has been used to model processes over time. Our analysis confirms that the evolving learning of social and individual features, as mobile phone sensing data accumulates over time, can also be fitted to the form of a *Gompertz* function. We see that this result is true for the prediction of different features, both social and individual, and for a set of different prediction methodologies, using a varying number of input signals, all collected via mobile phones in a field deployment.

Correlations between the evolution trends of the different learning process, as depicted in Figure 11, may imply underlying correlation between the raw data itself, and can hence be used as additional validation for correlated features and observations (such as the suggestion that people might have a higher tendency to marry within their own ethnic group, as has been widely observed [33,34]). In addition, this information could be used for informing the design of data collection configuration for an ongoing or future data collection initiative. For example, if we know of two features that are highly correlated in the same experiment, but one of them is very "cheap" to gather from a processing or battery power perspective, while the other is very expensive, we might decide that the cheaper one is sufficient (e.g. one requires just reading the phone's built-in call-log database while the other requires battery-intensive GPS scanning). Alternatively, we might want to make sure that two correlated values are gathered in order to strengthen the result and help deal with noise.

We can take our findings further, and extrapolate, using the learned Gompertz functions, the learning behavior and limits over time. Figure 12 shows the result of this extrapolation, two years into the future (the original source data is just one month). We can now gain different insights. First, extrapolation can be used to predict our maximal expected accuracy. In addition we can estimate where we are on each signal's estimated accuracy curve. We can then use this information to evaluate the analysis method, anticipate the timeline for increased accuracy, and understand when it is time to stop collecting/analyzing as we have reached a state of saturation. Another possible use is comparing different learning processes to one another, and using this information as part of the experiment or analysis management process. In addition, deviation from the expected curves might actually point at problems in the data collection process.

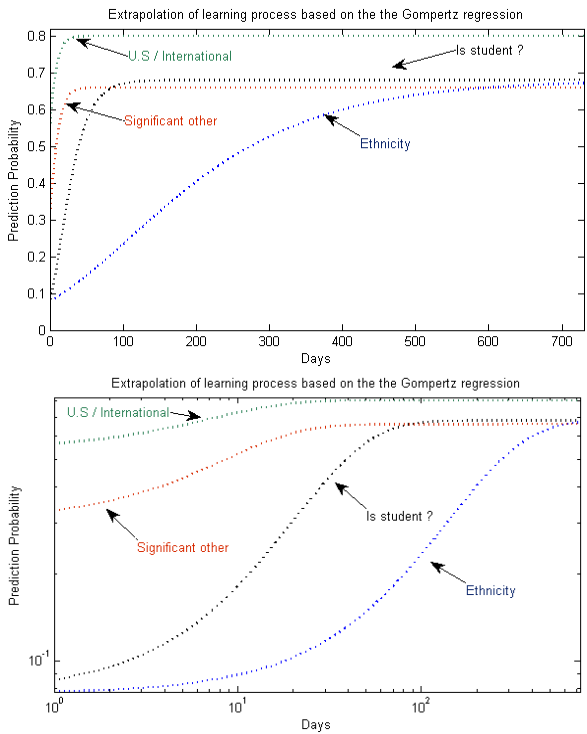

**Figure 12. Extrapolation of the learning process based on the Gompertz regression for the four learning tasks, in linear scale (top) and log-log scale (bottom).**

There are different reasons that might explain why there are the saturation limits in accuracy of our learning. For example, we can take the learning of user parameters which are based on Bluetooth proximity, as we have used in our prediction of significant-other ties and have also been used in [2,10, 28]. All of these analyses assume that the phone is an accurate proxy for its owner and is located where the owner is. It has been shown by Dey et al. [35] that people actually carry the phone with them much less than they might think. This discrepancy could account for some of the inaccuracies of trying to learn user parameters based on phone-sensed data.

Based on our observations, we can suggest this approach as a mechanism for answering several important questions, such as:

- Given a social network, how easy would it be for someone who monitors the behavioral activities of its members to infer it?
- What kinds of social features are more difficult to learn than others?
- What is the highest level of prediction accuracy that can be reached in a reasonable amount of time?

This could, in-turn, inform the allocation of data collection, processing, and analysis resources, as well as investigator time. Aside from their academic importance, such questions may also have significant financial implications. Social information has become a valuable data on its own merit, of high and tangible value, as it is used by many marketing and advertising platforms for doing targeted advertising to maximize their advertisement "hit" rates..

Furthermore, this insight may also have broader implications in areas of defense and homeland security, due to the importance of social information for cyber criminals and terrorists:

- Selling to highest bidder (both "legit" bidders, advertisers, etc., or in the black market to other attackers) [29] .
- Bootstrapping other attacks – e.g. using this as part of a complex *"Advanced Persistent Threats"* (APT) attack [30, 31].
- Business espionage - e.g. analyzing a competitor's customer base and profile high-yielding customers for targeted marketing [32].

## 6. CONCLUSIONS

The contributions of the work describes in this paper are the following:

- We demonstrated characteristics of incremental learning of multiple social and individual properties from raw sensing data collected from mobile phones, as the information is accumulated over time.
- We have shown that for different learning tasks, prediction methods, and input signals, the evolving learning of social and individual features, as mobile phone sensing data accumulates over time, can be fitted to the form of a *Gompertz* function.
- Furthermore, we proposed a method for advance prediction of the maximal learning accuracy possible for the learning task at hand, using just the first few measurements. This information can be useful in many ways, including:
  - Informing real-time resource allocation for data collection, for an ongoing data collection campaign.
  - Estimating accuracy limits and time needed for desired accuracy level of a given method.
  - Early evaluation of modeling and learning strategies.
- Finally, we presented new models for predicting social and individual features from raw mobile phone sensed data, which were developed as part of the methodological analysis.

Our main goal in this discussion was to investigate the learning process over time, rather than evaluate the specific models and how they generalize. In future work we intent to come back to each of these models and evaluate it in detail. We are also continuing our investigation of the properties of learning and prediction of human and social constructs based on mobile phone gathered data.

While there will always be the need for the expert and experienced "data artisan", with the exponential increase in accumulated data and the rise of a big-data ecosystem, there is an imperative need to create a more accurate science and engineering of data collection, processing, and analysis. Our work is a building block in this larger effort.